\documentclass[12pt,showpacs,aps]{revtex4}
\def\eu{\Uparrow_1\,}
\def\eeu{\Uparrow_2\,}
\def\ed{\Downarrow_1\,}
\def\eed{\Downarrow_2\,}
\def\nuu{\uparrow_1\,}
\def\nnu{\uparrow_2\,}
\def\nd{\downarrow_1\,}
\def\nnd{\downarrow_2\,}
\usepackage{graphicx}
\begin{document}
\title{Implementation of quantum logic operations \\ and creation of
entanglement\\ in a silicon-based quantum computer with constant interaction}
\author{G. P. Berman$^1$, G. W. Brown$^{2}$, M. E. Hawley$^{2}$,
D. I. Kamenev$^{1}$, V. I. Tsifrinovich$^{3}$}
\affiliation{$^1$Theoretical Division T-13,
Los Alamos National Laboratory, Los Alamos,
New Mexico 87545, USA}
\affiliation{$^2$Materials Science and Technology Division MST-8,
Los Alamos National Laboratory, Los Alamos,
New Mexico 87545, USA}
\affiliation{$^3$ IDS Department, Polytechnic University,
Brooklyn, New York 11201, USA}
\begin{abstract}
We describe how to implement quantum logic
operations in a silicon-based quantum computer with phosphorus atoms
serving as qubits. The information is stored in the states of nuclear spins and
the conditional logic operations are implemented through the electron
spins using nuclear-electron hyperfine and electron-electron exchange
interactions. The electrons in our computer should stay coherent
only during implementation of one Control-Not gate.
The exchange interaction is constant
and selective excitations are provided by a magnetic field gradient. 
The quantum logic operations are implemented by rectangular
radio-frequency pulses. This architecture is scalable
and does not require manufacturing nanoscale electronic gates.
As shown in this paper parameters of a quantum protocol can be
derived analytically even for
a computer with a large number of qubits using our perturbation approach.
We present the protocol for initialization of the nuclear spins and
the protocol for creation of entanglement. All analytical results are tested
numerically using a two-qubit system.
\end{abstract}
\maketitle

\section{Introduction}
The long decoherence time of nuclear spins of phosphorus donors in silicon
makes quantum computers based on these spins attractive
for quantum information processing. A scanning tunneling microscopy
technique~\cite{1994,1996,surfSci,electronics,clark}
can be used for creation of many identical
arrays of phosphorus atoms on the (100) surface of silicon. The phosphorus
qubits can be encapsulated by overgrowing additional silicon 
layers~\cite{clark}
to increase the electron relaxation time. Kane~\cite{kane} proposed to use
nanoscale electronic gates to control the qubits. This technique has not yet
been realized, due to fabrication issues,
and so in this paper we consider a different architecture. In our 
approach, the exchange interaction between qubits is constant and 
selective interactions are realized through the use of a magnetic 
field gradient and both microwave and radio frequency pulses.
Measurement can be implemented using optical
techniques~\cite{optical1,optical2,optical3}.
The measurement can be facilitated by creation of many
identical, noninteracting spin chains to amplify the signal.

\begin{figure}
%\vspace{-7mm}
\includegraphics[width=12cm,height=8cm]{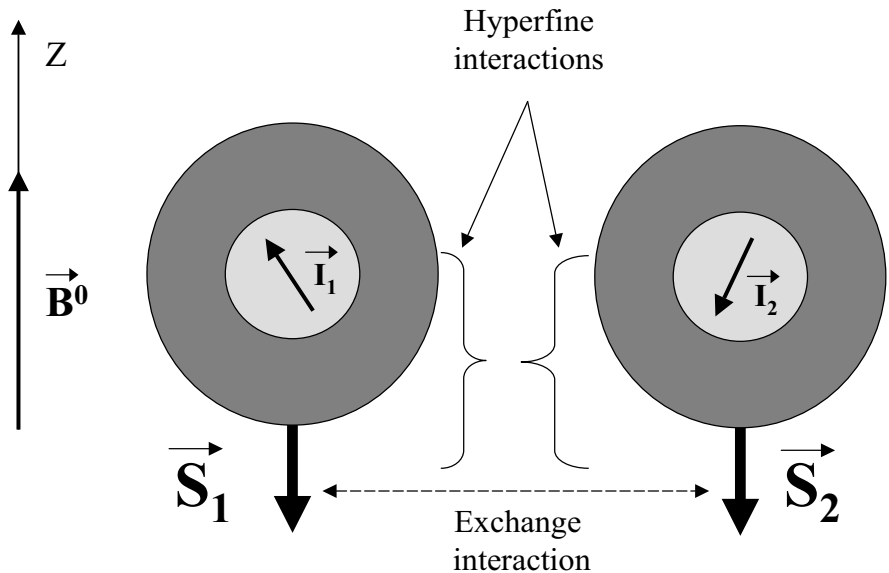}
\vspace{-4mm}
\caption{A schematic illustration of two phoshorus atoms placed
in a permanent magnetic field $\vec B^0$.
The electron spins $\vec S_1$ and
$\vec S_2$ (large arrows) of the neigboring atoms
interact with each other via the exchange interaction
and the nuclear spins $\vec I_1$ and
$\vec I_2$ (small arrows) interact with the electron spins
through the hyperfine interactions.}
\label{reffig:ldrd}
\end{figure}

The $^{31}$P atom has electron spin 1/2 and nuclear spin 1/2.
If the qubits in each chain are placed at a separation of $\sim$ 10 
nm from each other,
the nuclear-nuclear, nuclear-electron, and
electron-electron dipole-dipole interactions are small compared to the
electron-electron exchange interaction so that one can neglect
the dipole-dipole interactions (see Fig.~\ref{reffig:ldrd}).
There is also a relatively strong
hyperfine interaction between the electron and nuclear spins of
a $^{31}$P atom. Since the relaxation time
for the electron spins at temperatures of 1-7 K is relatively small
(0.6-60 ms at 7 K~\cite{T_c}), the quantum information
must be stored in the states of the nuclear spins.
Because the nuclear spins do not interact, electron spins can be used
to mediate the nuclear-nuclear interactions. In this setup,
the electron spins must be coherent only during the relatively short
time of implementation of a quantum logic gate, such as a Control-Not gate,
on a particular pair of qubits.

We consider in this paper a procedure for implementation
of entanglement between the nuclear spins in a two-qubit quantum
computer using rectangular radio-frequency pulses.
Entanglement is the simplest quantum logic operation
needed for implementation of more complex quantum logic gates
and is useful for demonstration of quantum computation in a
potentially scalable solid-state system. The paper is organized as follows.
The eigenstates of the system
are calculated analytically in Section II.
A brief description of the protocol for
creation of entanglement is given in Section III. The quantum
dynamics of the system is described in Section~IV.
The eigenstates from Section II are used for
calculation of pulse parameters in Section V.
Initialization and entanglement with two qubits
are simulated numerically in
Section VI. In Section VII we review the working conditions and
the parameter range for our computer.

\section{Eigenstates}
\label{sec:eigenstates}
The unperturbed Hamiltonian of the system reads
$$
\hat H^0=\gamma_e B_1^z\hat S_1^z+\gamma_e B_2^z\hat S_2^z-
\gamma_n B_1^z\hat I_1^z-\gamma_n B_2^z\hat I_2^z+
$$
$$
A\left[\hat S_1^z\hat I_1^z+\hat S_2^z\hat I_2^z+{1\over 2}\left(
\hat S_1^+\hat I_1^-+\hat S_1^-\hat I_1^++\hat S_2^+\hat I_2^-+
\hat S_2^-\hat I_2^+\right)\right]+
J\left[\hat S_1^z\hat S_2^z+{1\over 2}\left(
\hat S_1^+\hat S_2^-+\hat S_1^-\hat S_2^+\right)\right],
$$
where $\hat S_k^\alpha$ ($\hat I_k^\alpha$)
is the projection of the $k$th electron (nuclear) spin on the
$\alpha$th axis, $k=1,2$,
$\alpha=x,y,z$; $\hat S^{\pm}_k=\hat S^x_k\pm i\hat S^y_k$,
$\hat I^{\pm}_i=\hat I^x_k\pm i\hat I^y_k$;
$\gamma_e$ and $\gamma_n$ are, respectively, the electron and nuclear
gyromagnetic ratios; $B_k^z$ is the permanent magnetic field in the location
of the $k$th spin; $A$ and $J$ are, respectively, the hyperfine and exchange
interaction constants.

Let us define
$$
\hat \Sigma=\hat S_1^z+\hat S_2^z+\hat I_1^z+\hat I_2^z.
$$
The quantum states form 5 independent subspaces characterized by the
value of
$$
\Sigma_n=\langle\psi_n|\hat \Sigma|\psi_n\rangle.
$$
Two one-dimensional subspaces with $\Sigma_n=\pm 2$
are formed by the eigenstates with the following
eigenvalues $E_n$ and eigenvectors $|\psi_n\rangle$:
\begin{equation}
\label{E15}
E_{15}={1\over 2}\left(-\gamma_e B_1^z-\gamma_e B_2^z+
\gamma_n B_1^z+\gamma_n B_2^z\right)+{A\over 2}+{J\over 4}=
\left(-\gamma_e + \gamma_n \right)b+{A\over 2}+{J\over 4},
\end{equation}
\begin{equation}
\label{psi15}
|\psi_{15}\rangle=|\nnd\eed\ed\nd\rangle,
\end{equation}
\begin{equation}
\label{E0}
E_{0}=\left(\gamma_e - \gamma_n \right)b+{A\over 2}+{J\over 4},~~~
|\psi_{0}\rangle=|\nnu \eeu \eu \nuu\rangle,
\end{equation}
where $b=(B_2^z+B_1^z)/2$ and
we use the notation $|\Uparrow_k\rangle$ and
$|\Downarrow_k\rangle$ for the states of the electron spins and
$|\uparrow\rangle_k$ and $|\downarrow\rangle_k$ for the states
of the nuclear spins, $k=1,2$.

The basis vectors for the subspace with $\Sigma_n=-1$  are
\begin{equation}
\label{psiSigma-1}
\left(\begin{array}{c}
|\nnd \eed \eu \nd \rangle\\
|\nnd \eeu \ed \nd \rangle \\
|\nnd \eed \ed \nuu \rangle\\
|\nnu \eed \ed \nd \rangle\\
\end{array}\right)=
\left(\begin{array}{c}
| 13 \rangle\\
| 11 \rangle \\
| 14 \rangle\\
| 7 \rangle\\
\end{array}\right).
\end{equation}
Introducing the notation $\delta B=(B_2^z-B_1^z)/2$
the Hamiltonian matrix  for $\Sigma_n=-1$ becomes
\begin{equation}
\label{HSigma-1}
\left(\begin{array}{cccc}
-\gamma_e\delta B+\gamma_n b-{J\over 4} & {J\over 2} & {A\over 2} & 0 \\
{J\over 2} & \gamma_e\delta B+\gamma_n b-{J\over 4} & 0 & {A\over 2} \\
{A\over 2} & 0 & -\gamma_e b+\gamma_n\delta B+{J\over 4} & 0 \\
0 & {A\over 2} & 0 & -\gamma_e b-\gamma_n\delta B+{J\over 4}
\end{array}\right).
\end{equation}
This matrix can be diagonalized analytically.
For $\delta B=0$ the eigenvalues and
eigenfunctions were found in Ref.~\cite{2000}.
Instead of the exact analytical solution
we apply here a perturbative approach.
Our perturbative approach has an advantage over the exact
analytical solution because it can be
used to find the eigenstates for the quantum computer with more than
two qubits when no exact analytical solution is available.
The perturbation theory is based on the
fact that $\gamma_e b$ is at least three orders of magnitude larger than
$J/4$, $A/2$, $\gamma_e\delta B$, $\gamma_n b$, and
$\gamma_n\delta B$. (We take $b\approx 3.3$ T so that
$\gamma_e b/(2\pi)\approx 93$ GHz, $J/(2\pi)\approx 1-10$ MHz
or less, $A/(2\pi)=116$ MHz.) For our range of parameters the matrix
(\ref{HSigma-1}) splits into two relatively independent $2\times 2$
blocks~\cite{PRA01,JAM}.
The first block is formed by the matrix elements in the upper left corner
and the second block is formed in the lower right corner. The relative
independence of the two different blocks follows
from the facts that (a) the moduli of the differences between the
eigenvalues of each block are much smaller than
the moduli of the differences  ($\sim \gamma_e b$)
between the eigenvalues of the different blocks;
and (b) the matrix elements $A/2$ relating the different blocks
are much smaller than $\gamma_e b$.
The corrections to the wave function due to the neglected terms
are of the order of
\begin{equation}
\label{correction}
\epsilon={(A/2)\over \gamma_e b}\approx 6\times 10^{-4}
\end{equation}
and corrections $E^{(2)}_i$
to the energy levels $E^{(0)}_i$, $i=1,\dots, 14$,
are of the order of tens of kilohertz. These corrections are important
in order to flip the nuclear spins because the Rabi frequencies
of the nuclear spins are of the same magnitude.

The eigenvalues and the eigenfunctions are
\begin{equation}
\label{E7}
E_7^{(0)}=-\gamma_e b-\gamma_n \delta B+{J\over 4},~~~
|\psi_7^{(0)}\rangle=|\nnu \eed \ed \nd \rangle,
\end{equation}
\begin{equation}
\label{E14}
E_{14}^{(0)}=-\gamma_e b+\gamma_n \delta B+{J\over 4},~~~
|\psi_{14}^{(0)}\rangle=|\nnd \eed \ed \nuu \rangle,
\end{equation}
\begin{equation}
\label{E13}
E_{13}^{(0)}=
\gamma_n b-{J\over 4}-\sqrt{(\gamma_e \delta B)^2+{J^2\over 4}},
\end{equation}
\begin{equation}
\label{E11}
E_{11}^{(0)}=
\gamma_n b-{J\over 4}+\sqrt{(\gamma_e \delta B)^2+{J^2\over 4}}.
\end{equation}
The eigenfunctions corresponding to $E_{13}^{(0)}$ and $E_{14}^{(0)}$ are
\begin{equation}
\label{psi13}
|\psi_{13}^{(0)}\rangle=D_{13}|\nnd 
\rangle\otimes\left\{\left[\gamma_e \delta B +
\sqrt{(\gamma_e \delta B)^2+{J^2\over 4}}\right]|\eed\eu\rangle
-{J\over 2}|\eeu\ed\rangle\right\}\otimes|\nd \rangle,
\end{equation}
\begin{equation}
\label{psi11}
|\psi_{11}^{(0)}\rangle=D_{11}|\nnd \rangle\otimes\left\{{J\over 
2}|\eed\eu\rangle+
\left[\gamma_e \delta B +
\sqrt{(\gamma_e \delta B)^2+{J^2\over 4}}\right]
|\eeu\ed\rangle\right\}\otimes |\nd\rangle,
\end{equation}
where $D_i$, $i=1,\dots,14$, are the normalization constants.
The correction $E^{(2)}_7$ is calculated in Appendix A and
all corrections $E^{(2)}_i$, $i=1,14$, are listed in Appendix B.

The basis vectors for the subspace with $\Sigma_n=1$  are
$$
\left(\begin{array}{c}
|\nnu \eeu \ed \nuu \rangle \\
|\nnu \eed \eu \nuu \rangle \\
|\nnu \eeu \eu \nd \rangle \\
|\nnd \eeu \eu \nuu \rangle \\
\end{array}\right)=
\left(\begin{array}{c}
| 2 \rangle\\
| 4 \rangle \\
| 1 \rangle\\
| 8 \rangle\\
\end{array}\right).
$$
The Hamiltonian matrix  for $\Sigma_n=1$ has the following form:
\begin{equation}
\label{HSigma1}
\left(\begin{array}{cccc}
\gamma_e\delta B-\gamma_n b-{J\over 4} & {J\over 2} & {A\over 2} & 0 \\
{J\over 2} & -\gamma_e\delta B-\gamma_n b-{J\over 4} & 0 & {A\over 2} \\
{A\over 2} & 0 & \gamma_e b-\gamma_n\delta B+{J\over 4} & 0 \\
0 & {A\over 2} & 0 & \gamma_e b+\gamma_n\delta B+{J\over 4}
\end{array}\right).
\end{equation}

The eigenvalues and the eigenfunctions are
\begin{equation}
\label{E1}
E_1^{(0)}=\gamma_e b-\gamma_n \delta B+{J\over 4},~~~
|\psi_1^{(0)}\rangle=|\nnu \eeu \eu \nd \rangle,
\end{equation}
\begin{equation}
\label{E8}
E_{8}^{(0)}=\gamma_e b+\gamma_n \delta B+{J\over 4},~~~
|\psi_{8}^{(0)}\rangle=|\nnd \eeu \eu \nuu \rangle,
\end{equation}
\begin{equation}
\label{E2}
E_{2}^{(0)}=
-\gamma_n b-{J\over 4}+\sqrt{(\gamma_e \delta B)^2+{J^2\over 4}},
\end{equation}
\begin{equation}
\label{E4}
E_{4}^{(0)}=
-\gamma_n b-{J\over 4}-\sqrt{(\gamma_e \delta B)^2+{J^2\over 4}},
\end{equation}
\begin{equation}
\label{psi2}
|\psi_{2}^{(0)}\rangle=D_2|\nnu\rangle\otimes\left\{\left[\gamma_e \delta B +
\sqrt{(\gamma_e \delta B)^2+{J^2\over 4}}\right]
|\eeu\ed\rangle+{J\over 2}
|\eed\eu\rangle
\right\}\otimes|\nuu\rangle,
\end{equation}
\begin{equation}
\label{psi4}
|\psi_{4}^{(0)}\rangle=D_4|\nnu\rangle\otimes\left\{-{J\over 2}|\eeu\ed\rangle+
\left[\gamma_e \delta B +
\sqrt{(\gamma_e \delta B)^2+{J^2\over 4}}\right]|\eed\eu\rangle
\right\}\otimes|\nuu\rangle.
\end{equation}

The six-dimensional multiplet with $\Sigma_n=0$ splits into three relatively
independent subspaces. The first two eigenvalues and eigenfunctions are
\begin{equation}
\label{E6}
E_6^{(0)}=-\gamma_e b-\gamma_nb-{A\over 2}+{J\over 4},~~~
|\psi_6^{(0)}\rangle=|\nnu\eed\ed\nuu\rangle,
\end{equation}
\begin{equation}
\label{E9}
E_{9}^{(0)}=\gamma_e b+\gamma_nb-{A\over 2}+{J\over 4},~~~
|\psi_9^{(0)}\rangle=|\nnd\eeu\eu\nd\rangle.
\end{equation}
The remaining four eigenstates with the basis vectors
$$
\left(\begin{array}{c}
|\nnu \eeu \ed \nd \rangle \\
|\nnu \eed \eu \nd \rangle \\
|\nnd \eeu \ed \nuu \rangle \\
|\nnd \eed \eu \nuu \rangle \\
\end{array}\right)=
\left(\begin{array}{c}
| 3 \rangle\\
| 5 \rangle \\
| 10 \rangle\\
| 12 \rangle\\
\end{array}\right)
$$
are related to the following Hamiltonian matrix:
%\footnotesize
\small
\begin{equation}
\label{HSigma0}
\left(\begin{array}{cccc}
\delta B(\gamma_e-\gamma_n)+{A\over 2}-{J\over 4} &
{J\over 2} & 0 & 0 \\
{J\over 2} & \delta B(-\gamma_e-\gamma_n)-{A\over 2}-{J\over 4}
& 0 & 0 \\
0 & 0 &
\delta B(\gamma_e+\gamma_n)-{A\over 2}-{J\over 4} & {J\over 2} \\
0 & 0 & {J\over 2} &
\delta B(-\gamma_e+\gamma_n)+{A\over 2}-{J\over 4}
\end{array}\right).
\end{equation}
\normalsize
One can see that these states form two independent (in our approximation)
two-dimensional subspaces. The first subspace is described by the
$2\times 2$ matrix in the upper left corner and the second subspace
is described by the $2\times 2$ matrix in the lower right corner of
the matrix~(\ref{HSigma0}). The eigenvalues and eigenfunctions are
\begin{equation}
\label{E3}
E_{3}^{(0)}=-\gamma_n \delta B-{J\over 4}+
\sqrt{\left(\gamma_e \delta B+{A\over 2}\right)^2+{J^2\over 4}},
\end{equation}
\begin{equation}
\label{psi3}
|\psi_{3}^{(0)}\rangle=D_3|\nnu\rangle\otimes\left\{\left[\gamma_e \delta B
+{A\over 2}+
\sqrt{\left(\gamma_e \delta B+{A\over 2}\right)^2+{J^2\over 4}}\right]
|\eeu\ed\rangle+{J\over 2}
|\eed\eu\rangle
\right\}\otimes|\nd\rangle,
\end{equation}
\begin{equation}
\label{E5}
E_{5}^{(0)}=-\gamma_n \delta B-{J\over 4}-
\sqrt{\left(\gamma_e \delta B+{A\over 2}\right)^2+{J^2\over 4}},
\end{equation}
\begin{equation}
\label{psi5}
|\psi_{5}^{(0)}\rangle=D_5|\nnu \rangle\otimes\left\{-{J\over 
2}|\eeu\ed\rangle+
\left[\gamma_e \delta B + {A\over 2} +
\sqrt{\left(\gamma_e \delta B+ {A\over 2}\right)^2+
{J^2\over 4}}\right]|\eed\eu\rangle
\right\}\otimes|\nd\rangle,
\end{equation}
\begin{equation}
\label{E10}
E_{10}^{(0)}=\gamma_n \delta B-{J\over 4}-
\sqrt{\left(\gamma_e \delta B-{A\over 2}\right)^2+{J^2\over 4}},
\end{equation}
\begin{equation}
\label{psi10}
|\psi_{10}^{(0)}\rangle=D_{10}|\nnd\rangle\otimes\left\{-{J\over 
2}|\eeu\ed\rangle+
\left[\gamma_e \delta B - {A\over 2} +
\sqrt{\left(\gamma_e \delta B - {A\over 2}\right)^2+
{J^2\over 4}}\right]|\eed\eu\rangle
\right\}\otimes|\nuu\rangle,
\end{equation}
\begin{equation}
\label{E12}
E_{12}^{(0)}=\gamma_n \delta B-{J\over 4}+
\sqrt{\left(\gamma_e \delta B-{A\over 2}\right)^2+{J^2\over 4}},
\end{equation}
\begin{equation}
\label{psi12}
|\psi_{12}^{(0)}\rangle=D_{12}|\nnd 
\rangle\otimes\left\{\left[\gamma_e \delta B -
{A\over 2}+
\sqrt{\left(\gamma_e \delta B-{A\over 2}\right)^2+{J^2\over 4}}\right]
|\eeu\ed\rangle+{J\over 2}
|\eed\eu\rangle
\right\}\otimes|\nuu\rangle.
\end{equation}
The eigenvalues $E_{10}$ and $E_{12}$ in Eqs.~(\ref{E10}) and (\ref{E12})
are written for the case $A/2\ge \gamma_e\delta B$. For the opposite case,
$A/2< \gamma_e\delta B$, one must exchange the eigenvalues
$E_{10}\leftrightarrow E_{12}$ and leave the eigenvectors $|\psi_{10}\rangle$
and $|\psi_{12}\rangle$ unchanged.

\section{Creation of entanglement}
\label{sec:entanglement}
Consider the quantum dynamics generated by electromagnetic
pulses for different parameters $J$ and $\delta B$.
The four states, 6th, 7th, 14th, and 15th, have the lowest energies
of the order of $-\gamma_e b$. The distance
between the lower 6th and the upper 15th levels of the quartet
is $(2\gamma_n b+A)/(2\pi)\approx 173$ MHz
[$\gamma_n/(2\pi)\approx 17.25144$ MHz/T]. This is much smaller than
$k_{\rm B}T=20.83$ GHz, where $T=1$ K is the temperature.
Consequently, all four of these states are initially populated.

The initialization of the nuclear spins and creation of
entanglement between the nuclear spins can be implemented by using the
fact that the electron spins are polarized. Our system
can be represented as a one-dimensional spin chain
\begin{equation}
\label{basis}
|n_2e_2e_1n_1\rangle.
\end{equation}
In Eq.~(\ref{basis}) $e_1$ assumes the values $\eu, \ed$, $e_2=\eeu,\eed$,
$n_1=\nuu,\nd$, and $n_2=\nnu,\nnd$,
In the spin chain (\ref{basis})
there are interactions only between the neighboring spins,
so that this kind of spin ordering is convenient
for analysis of conditional quantum logic gates.

Initially our chain is in the superposition of states with the lowest
energies $E_6$, $E_7$, $E_{14}$, and $E_{15}$.
 From Eqs.~(\ref{psi15}), (\ref{E7}), (\ref{E14}), and (\ref{E6})  one can see
that these states are
\begin{equation}
\label{initial0}
|n_2\eed \ed n_1\rangle,
\end{equation}
with different $n_1$ and $n_2$. One possible initial state is shown
in Fig.~\ref{reffig:ldrd}. By using Control-Not gates
between the electron and nuclear spins, one can transfer the polarization
from electron to nuclear spins. After some time the electron spins
polarize and one obtains the only populated state
$$
|\nnd\eed \ed \nd\rangle.
$$
By using the Hadamard transform on the 1st nuclear spin,
Control-Not gate between the 1st nuclear spin and 1st electron spin,
Control-Not gate between the 1st electron spin and 2nd electron spin,
and Control-Not gate between the 2nd electron spin and 2nd nuclear spin
one can create entanglement between all spins of the system
\begin{equation}
\label{entangle}
{1\over\sqrt 2}\left(|\nnu\eeu \eu \nuu\rangle+e^{i\theta}
|\nnd\eed \ed \nd\rangle\right).
\end{equation}
The exact value of the phase $\theta$ is not important for us.

\section{Dynamics}
\begin{figure}
%\vspace{-5mm}
\includegraphics[width=14cm,height=6cm]{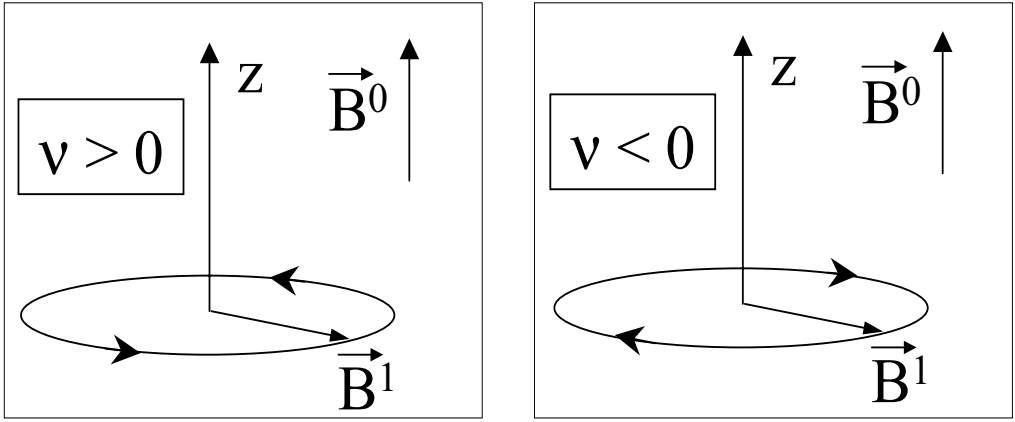}
\vspace{-3mm}
\caption{Different kinds of polarization of the electromagnetic wave for
$\nu>0$ and $\nu<0$.}
\label{reffig:nu}
\end{figure}

The time-dependent magnetic field has the following components:
\begin{equation}
\label{B1t}
\vec B^1(t)=B^1(\cos(\nu t+\varphi),-\sin(\nu t+\varphi),0),
\end{equation}
where $B^1$, $\nu$, and $\varphi$ are, respectively, amplitude, frequency
and phase of the pulse and $t$ is time. The frequency
$\nu$ can assume both positive and negative values
as shown in Fig.~\ref{reffig:nu}.
The perturbation term in the Hamiltonian has the form
$$
\hat V(t)=\left[{\Omega_e^0\over 2}\left(\hat S_1^-+\hat S_2^-\right)-
{\Omega_n^0\over 2}\left(\hat I_1^-+\hat I_2^-\right)\right]e^{-i(\nu 
t+\varphi)}+
$$
\begin{equation}
\label{Vt}
\left[{\Omega_e^0\over 2}\left(\hat S_1^++\hat S_2^+\right)-
{\Omega_n^0\over 2}\left(\hat I_1^++\hat I_2^+\right)\right]e^{i(\nu 
t+\varphi)},
\end{equation}
where $\Omega_e^0=\gamma_e B^1$ and $\Omega_n^0=\gamma_n B^1$.

\subsection{Scheme for numerical simulations}
\label{sec:numerical1}
The numerical simulations are performed without using the perturbation
approach and results are presented in Sec.~\ref{sec:numerical} below.
It is convenient to work in the
rotating frame where the effective Hamiltonian is independent of time.
The relationship between the wave function $|\Psi(t)\rangle$ in the
laboratory frame and the wave function $|\Phi(t)\rangle$ in the
rotating frame is
\begin{equation}
\label{labRot}
|\Psi(t)\rangle=e^{i\hat\Sigma(\nu t +\varphi)}|\Phi(t)\rangle.
\end{equation}
The Schr\"odinger equation in the rotating frame is
\begin{equation}
\label{SchPhi}
i|\dot \Phi(t)\rangle=
\left(\hat H^0+\nu\hat \Sigma+\hat V\right)|\Phi(t)\rangle,
\end{equation}
\begin{equation}
\label{V}
\hat V=\Omega_e(\hat S^x_1+\hat S^x_2)-\Omega_n^0(\hat I^x_1+\hat I^x_2).
\end{equation}
Let us decompose the wave function over the eigenstates $|\psi_n\rangle$
of the Hamiltonian $\hat H_0$ as
\begin{equation}
\label{decompPhi}
|\Phi(t)\rangle=\sum_{n=0}^{15}c_n(t)|\psi_n\rangle,
\end{equation}
where the functions $|\psi_n\rangle$ are related to the basis
functions $|i\rangle$ by
\begin{equation}
\label{a_ni}
|\psi_n\rangle=\sum_ia_{n,i}|i\rangle.
\end{equation}
The coefficients $a_{n,i}$ are calculated in Section~\ref{sec:eigenstates}
in the zeroth order approximation.
(For our numerical simulations we use the exact values of $a_{n,i}$.)
The system of 16 differential equations for the expansion
coefficients $c_n(t)$ is
\begin{equation}
\label{dot_c_n}
i\dot c_n(t)=\left(E_n+\Sigma_n\nu\right)c_n(t)+\sum_{m=0}^{15}V_{nm}c_m(t),
\end{equation}
where
\begin{equation}
\label{Vnm}
V_{n,m}=\langle\psi_n|\hat V|\psi_m\rangle.
\end{equation}
Eq.~(\ref{dot_c_n}) can be regarded as the Schr\"odinger equation
\begin{equation}
\label{dot_ct}
i\dot c_n(t)=\sum_{m=0}^{15} h_{n,m}c_m(t)
\end{equation}
with the time-independent Hamiltonian $\hat h$ whose matrix elements
have the following form:
\begin{equation}
\label{h_nm}
h_{n,m}=(E_n+\nu \Sigma_n)\delta_{nm}+V_{nm}.
\end{equation}
The dynamics of the coefficients $c_n(t)$ can be computed using
the eigenfunctions $b^q_n$, $q=0,\dots,15$ and the eigenvalues $e^q$
of the Hamiltonian $\hat h$ as
\begin{equation}
\label{dynamics_rotating}
c_n(t)=\sum_{m=0}^{15}c_m(t_0)\sum_{q=0}^{15}\left(b^q_m\right)^*b^q_n
e^{-ie^q(t-t^\prime)},
\end{equation}
where $t^\prime$ is the time of the beginning of the pulse.
The wave function in the laboratory frame can be represented as
\begin{equation}
\label{decompPsi}
|\Psi(t)\rangle=\sum_{p=0}^{15}D_p(t)e^{-iE_pt}|\psi_p\rangle.
\end{equation}
Before each pulse at time $t=t^\prime$ we make the transformation
to the rotating frame using the relation
[see Eqs.~(\ref{labRot}), (\ref{decompPhi}) and (\ref{decompPsi})]
\begin{equation}
\label{Ac}
D_p(t)=\exp\left[iE_pt+i\Sigma_p(\nu t+\varphi)\right]c_p(t),
\end{equation}
and after the pulse at the time $t=t^\prime+\tau$
($\tau$ is the duration of the pulse) we make the back transformation
to the laboratory frame using the same formula.

\section{Implementation of logic gates}
We describe here implementation of logic gates in terms of
the basis states $|i\rangle$, $i=0,\dots,15$. From
Section~\ref{sec:eigenstates} the basis functions $|i\rangle$
are approximately equal to the eigenfunctions $|\psi_i\rangle$
if the conditions $\gamma_e\delta B\gg J/2$ and
$|\gamma_e\delta B-A/2|\gg J/2$ are satisfied.
Assume that the frequency of a pulse is close
to the transition frequency of the $k$th ($k=1,2$) electron
or nuclear spin; the direction of the spin in the state $|p\rangle$
is along the direction of the permanent magnetic field $\vec B^0$ (i.e.
$|p\rangle=|\dots \uparrow_k\dots\rangle$ or
$|p\rangle=|\dots \Uparrow_k\dots\rangle$);
and the state $|q\rangle$ is related to the state $|p\rangle$ by a
flip of the $k$th spin. For the initial conditions
$$
D_p(t^\prime)=1,\qquad D_q(t^\prime)=0
$$
the dynamics of this spin is described by the following
equations~\cite{book,book1}:
$$
D_p(t^\prime+\tau)=\left\{\cos\left[{\lambda_{q,p}\tau\over 2}\right]+
i{\Delta_{q,p}\over\lambda_{q,p}}
\sin\left[{\lambda_{q,p}\tau\over 2}\right]\right\}
e^{-i\Delta_{q,p}\tau/2},
$$
\begin{equation}
\label{solution}
D_q(t^\prime+\tau)=i{\Omega\over\lambda_{q,p}}
\sin\left[{\lambda_{q,p}\tau\over 2}\right]
e^{i\Delta_{q,p} t^\prime-i\varphi+i\Delta_{q,p}\tau/2},
\end{equation}
where
$$
\Delta_{q,p}=E_q-E_p-\nu,~~ \lambda_{q,p}=\sqrt{\Delta^2_{q,p}+\Omega^2},
$$
$t^\prime$ is the time of the beginning of the pulse, $\tau$ is the 
duration of the pulse, $\Omega=\Omega_e$ is the Rabi frequency
of the electron spin, and $\Omega=\Omega_n$  is the Rabi frequency
of the nuclear spin. For the other initial conditions
$$
D_p(t^\prime)=0,\qquad D_q(t^\prime)=1
$$
the solution is
$$
D_p(t^\prime+\tau)=i{\Omega\over\lambda_{q,p}}
\sin\left[{\lambda_{q,p}\tau\over 2}\right]
e^{-i\Delta_{q,p} t^\prime+i\varphi-i\Delta_{q,p}\tau/2},
$$
\begin{equation}
\label{solution1}
D_q(t^\prime+\tau)=\left\{\cos\left[{\lambda_{q,p}\tau\over 2}\right]-
i{\Delta_{q,p}\over\lambda_{q,p}}
\sin\left[{\lambda_{q,p}\tau\over 2}\right]\right\}
e^{i\Delta_{q,p}\tau/2}.
\end{equation}
The complete transition between the states $|p\rangle$ and $|q\rangle$
takes place when the detuning is equal to $\Delta_{q,p}=0$ and when
$\tau =\pi/\Omega$ ($\pi$-pulse). A  near-resonant transition
with $\Delta\ne 0$ can be completely suppressed when the
condition~\cite{book}
\begin{equation}
\label{2piK}
\Omega={|\Delta_{q,p}|\over \sqrt{4K^2-1}},
\end{equation}
known as the $2\pi K$ condition, is satisfied.
Here $K=1, 2,\dots$ is an integer number. For this value
of $\Omega$ the value of the sine in Eqs.~(\ref{solution}) and 
(\ref{solution1})
is equal to zero.

As follows from the above considerations, (i) in order to implement a
complete transition, the frequency of the pulse must be resonant
to this transition, $\nu=E_q-E_p$, where $\nu$
can assume both positive and
negative values. (ii) In order to completely suppress a transition with
$\Delta\ne 0$, the Rabi frequency of the pulse must satisfy the
$2\pi K$-condition~(\ref{2piK}). Both operations (i) and (ii) can be 
implemented
simultaneously by one pulse if there are two states in the quantum
register. Actually, we described here the procedure for
implementation of the Control-Not gate,
which we will use to create entanglement in our system.

We now derive the parameters of the gates described in
Section~\ref{sec:entanglement}.
Let CN$_{l,k}$ be the Control-Not gate which
flips the $k$th spin in the state $|i\rangle$
and suppresses the flip of the same spin in the state $|j\rangle$
(the latter has different orientation of the $l$th spin).
We assume that the $k$th
spin is pointed up (along the direction of $\vec B^0$)
in both states $|i\rangle$ and $|j\rangle$.
Let the state $|i^\prime\rangle$ be related to the state
$|i\rangle$ by the flip of the
$k$th spin and the state $|j^\prime\rangle$ be related to the state
$|j\rangle$ by the flip of the same spin. Then the frequency $\nu$ and the
detuning $\Delta_{j^\prime,j}$ in Eq.~(\ref{2piK}) for the Rabi frequency are
\begin{equation}
\label{nuOmega}
\nu=E_{i^\prime}-E_{i},~~~
\Delta_{j^\prime,j}=E_{j^\prime}-E_{j}-\nu.
\end{equation}
In our notation, it is convenient to treat the state of the electron spin
$|\Uparrow_k\rangle$ as $|0_k\rangle$ and the state
$|\Downarrow_k\rangle$ as $|1_k\rangle$. The fact that
the energy of the state $|0_k\rangle$ is larger than the energy of the state
$|1_k\rangle$ is accounted for by a negative frequency $\nu$ of the pulse
in the first equation~(\ref{nuOmega}) .

\subsection{Initialization}
The nuclear spins can be polarized by using the fact that
the Larmor frequencies of the electron spins
depend on orientations of the corresponding nuclear spins through the
hyperfine interaction. Measuring the electron Larmor frequencies using,
for example, a scanning tunneling microscope~\cite{init1,init2}, one can
define the orientation of the nuclear spins and apply a selective
$\pi$-pulse if necessary. Here we describe a different technique.
We assume that initially the nuclear spins are not polarized, the
electron spins are polarized, and there are four states
(\ref{initial0}) in the register. If we apply the gates
\begin{equation}
\label{init1}
{\rm CN}_{e_1,n_1}{\rm CN}_{n_1,e_1}
\end{equation}
(the order of implementation of the operators is from the right to the left),
we obtain the superposition of states
\begin{equation}
\label{states1}
|n_2\eed e_1\nd\rangle
\end{equation}
with indefinite orientation of the first electron spin and with definite
orientation of the first nuclear spin.

Since the state of the first electron spin in Eq.~(\ref{states1})
is unknown, we cannot immediately swap the states of the second
electron and nuclear spins because of the interaction between the electron
spins. One has to wait while the electron spin polarizes again
(during, for example, the time-interval 0.1 s).
In our numerical simulations the relaxation
of the electron spins is modeled by flipping them ``by hand'', without using
electromagnetic pulses, in all states of superposition.

After the electron spins are polarized one applies the gates
\begin{equation}
\label{initGates}
{\rm CN}_{e_2,n_2}{\rm CN}_{n_2,e_2}
\end{equation}
and obtains the state
\begin{equation}
\label{states2}
|\nnd e_2 \ed\nd\rangle.
\end{equation}
One waits while the second electron spin relaxes and obtains the state
\begin{equation}
\label{states3}
|\nnd\eed\ed\nd\rangle
\end{equation}
which is used as an initial state for creation of the entanglement.

\subsection{Entanglement}
The sequence of gates
\begin{equation}
\label{entanglementGates}
\overline{{\rm CN}}_{e_2,n_2}\overline{{\rm CN}}_{e_1,e_2}
\overline{{\rm CN}}_{n_1,e_1}{\rm Had}_{n_1}
\end{equation}
generates the entangled state (\ref{entangle}).
In Eq.~(\ref{entanglementGates})
${\rm Had}_{n_1}$ is the Hadamard gate on the first nuclear spin,
and the gate $\overline{{\rm CN}}_{k,k^\prime}$ is the inverse of the
Control-Not gate: it flips the target qubit only if the control qubit
is in the state $|0_k\rangle$. The parameters of these gates
can be calculated analytically using Eq.~(\ref{nuOmega}).
In our simulations presented below, the Hadamard gate is
performed by applying a $\pi/2$ pulse of duration
$\tau=\pi/(2\Omega_n)$, where $\Omega_n$ is given by
Eq.~(\ref{Omega_tau_n}) below.

In principle, it is possible to inplement a Control-Not gate 
between the nuclear spins without changing the states of the 
electron spins~\cite{adiabatic1}. In practice, this approach is not useful
because one can show (using calculated in this paper eigenvalues) that
the mediated by the electrons effective coupling between the 
nuclear spins is of the order of 3.5 Hz or less. This means that 
the Rabi frequency of the pulse implementing the Control-Not gate 
must be less than 2 Hz and the frequency of the pulse must be tuned in
resonance with the accuracy of approximately 0.1 Hz.  
 
\subsection{Rabi frequencies}
\label{sec:Rabi}
The Rabi frequency of the electron spins $\Omega_e$
is different from $\Omega_e^0=\gamma_e B^1$ because of the
exchange interaction between the electron spins. Similarly,
the Rabi frequency of the nuclear spins $\Omega_n$
is different from $\Omega_n^0=\gamma_n B^1$ because of the
hyperfine interaction between the nuclear and electron spins.

Consider, for example, the transition
\begin{equation}
\label{initTransition}
|\nnu\eed \ed \nuu\rangle\rightarrow |\nnu\eed \eu \nuu\rangle,
\end{equation}
associated with the flip of the first electron spin. The matrix
element $V_{6,4}=\Omega_e/2$ of the matrix $h_{n,m}$ in Eq.~(\ref{h_nm})
is responsible for this transition. The value
of $\Omega_e$ must satisfy the $2\pi K$ condition~(\ref{2piK}).
The value of  $V_{6,4}$ can be calculated using Eq.~(\ref{Vnm}). Only
two terms in the sum
\begin{equation}
\label{Vnm_sum}
V_{nm}=\sum_{i,j=0}^{15}a_{i,n}\langle i|\hat V|j\rangle a_{m,j},
\end{equation}
where $n=6,~m=4$, appreciably contribute.  These are
\begin{equation}
\label{V46}
V_{6,4}=\sum_{i,j=0}^{15}a_{i,6}\langle i|\hat V|j\rangle a_{4,j}
\approx a_{6,6}\langle 6|\hat V|4\rangle a_{4,4}+
a_{6,6}\langle 6|\hat V|2\rangle a_{4,2}.
\end{equation}
The first term is due to the flip of the first electron spin by the 
electromagnetic pulse, i.e., due to the transition
\begin{equation}
\label{transition20}
|\nnu\eed \ed \nuu\rangle\rightarrow |\nnu\eed \eu \nuu\rangle,
\end{equation}
The second term is due to the two-step transition
\begin{equation}
\label{transition21}
|\nnu\eed \ed \nuu\rangle\rightarrow |\nnu\eeu \ed \nuu\rangle
\rightarrow |\nnu\eed \eu \nuu\rangle,
\end{equation}
where the first step is due to the
nonselective excitation of the second electron spin
and the second step is implemented due to the exchange interaction
between the electrons. The contribution of the second term is
proportional to the ratio of the matrix element $J/2$, responsible for the
exchange interaction, [see the second transition in Eq.~(\ref{transition21})]
to the detuning $2\delta B$, characterizing the first transition.
 From Eq.~(\ref{V46}) the Rabi frequency is
\begin{equation}
\label{Omega_e}
\Omega_e\approx \Omega_e^0\left(1-{J\over 4\gamma_e\delta B}\right).
\end{equation}
This expression for the electron Rabi frequency holds for all other electron
transitions.
The $2\pi K$ condition (\ref{2piK}) for the electron spin becomes
\begin{equation}
\label{2piKe}
\Omega^0_e=\gamma_e B^1={|\Delta_{j^\prime,j}|\over \sqrt{4K_e^2-1}}
\left(1-{J\over 4\gamma_e \delta B}\right)^{-1},
\end{equation}
where $|\Delta_{j^\prime,j}|\approx A$ when the control spin is the
nuclear spin and $|\Delta_{j^\prime,j}|\approx J$ when the control spin is the
neighboring electron spin.

We now find the Rabi frequency of the nuclear spin.
Consider the gate ${\rm CN}_{e_1,n_1}$ in Eq.~(\ref{init1})
acting on the state $|\nnu\eed \ed \nuu\rangle$.
The nuclear spin is flipped as a result of the transition
\begin{equation}
\label{initTransition1}
|\nnu\eed \ed \nuu\rangle\rightarrow |\nnu\eed \ed \nd\rangle,
\end{equation}
which is implemented through the matrix
element $V_{6,7}=-\Omega_n/2$. In Eq.~(\ref{Vnm}), only
two terms considerably contribute to the value of $V_{6,7}$, namely
\begin{equation}
\label{V67}
V_{6,7}\approx a_{6,6}\left(-{\Omega_n^0\over 2}\right)a_{7,7}+
a_{5,6}{\Omega_e^0\over 2}a_{7,7}.
\end{equation}
The first term is due to the flip of the nuclear spin by the 
electromagnetic pulse in Eq.~(\ref{initTransition1}).
The second term is due to the two-step transition
\begin{equation}
\label{transition11}
|\nnu\eed \ed \nuu\rangle\rightarrow |\nnu\eed \eu \nd\rangle
\rightarrow |\nnu\eed \ed \nd\rangle,
\end{equation}
where the first step is due to the hyperfine interaction between
the first electron spin and the first nuclear spin
and the second step is due to the
nonresonant action of the electromagnetic pulse
on the first electron spin.
The transition~(\ref{transition11}) is initiated
by the electron spin, which creates
a magnetic field in the $x$ direction comparable to the
magnetic field $B^1$ of the pulse.
In spite of the fact that the probability $P_e$ of flipping the electron spin
is small, the fact that
$\Omega_e/\Omega_n^0\sim 10 ^{3}$ makes the probability
of the transition (\ref{transition11}) comparable to the probability
of the transition~(\ref{initTransition1}).

In order to calculate $a_{5,6}$ in Eq.~(\ref{V67}), we must calculate the
first-order correction to the wave function~(\ref{psi5}) using the equation
\begin{equation}
\label{psi_correction}
|\psi_n^{(1)}\rangle=\sum_{m=0\atop m\ne n}^{15}
{V_{n,m}\over E^{(0)}_m-E^{(0)}_n}|\psi_m^{(0)}\rangle.
\end{equation}
Taking $n=5$ and multiplying by $\langle 6|$ we obtain
\begin{equation}
\label{psi_correction1}
a_{5,6}=\sum_{m=0\atop m\ne 5}^{15}
{V_{5,m}a_{m,6}\over E^{(0)}_m-E^{(0)}_5}.
\end{equation}
Only one term with $m=6$ significantly contributes to the sum, where
$$
V_{5,6}\approx\langle 5|\hat V|6\rangle={A\over 2}.
$$
We find
$$
a_{5,6}\approx
{A\over 2[E^{(0)}_6-E^{(0)}_5]}\approx -{A\over 2\gamma_e b}\ll 1.
$$
Putting this value to Eq.~(\ref{V67}) we obtain
the expression for the Rabi frequency of the nucleus
\begin{equation}
\label{Omega_correction}
\Omega_n=\Omega_n^0+{A\over 2\gamma_e b}\Omega_e^0\approx
2\Omega_n^0.
\end{equation}
This equation is valid for the Rabi frequencies associated with other
nuclear spin transitions.
Similar to Eq.~(\ref{Omega_e}) for the electron spin, the correction
to the Rabi frequency is proportional to the ratio
of the matrix element $A/2$, responsible for the hyperfine interaction,
to the detuning $\gamma_e b\approx\gamma_e b-\gamma_n b$ between
the frequencies of the nuclear and electron spins.

The $2\pi K$ condition for the nuclear spin reads
\begin{equation}
\label{2piKn}
\Omega^0_n=\gamma_n B^1={|\Delta_{j^\prime,j}|\over \sqrt{4K_n^2-1}}
\left(1+{A\over 2\gamma_n b}\right)^{-1},
\end{equation}
where $|\Delta_{j^\prime,j}|\approx A$.
We have calculated the Rabi frequencies only for the two gates.
The Rabi frequencies for the other gates can be calculated using the same
formulas~(\ref{2piKe}) and (\ref{2piKn}).

We still have indefinite parameters $K_e$ and $K_n$ in
Eqs.~(\ref{2piKe}) and (\ref{2piKn}) for the Rabi frequencies
of electron and nuclear spins. Increasing $K_e$ and $K_n$
decrease the electron and nuclear Rabi frequencies
which should satisfy the conditions
\begin{equation}
\label{smallOmega}
\varepsilon_e={\Omega_e\over 4\gamma_e\delta B}\ll 1,~~~
\varepsilon_n={\Omega_n\over 4\gamma_n\delta B}\ll 1.
\end{equation}
These conditions provide selective excitations of the
spins~\cite{nonresonant}. The error in the probability amplitude
due to nonselective excitations is of the order of
$\varepsilon_e$ for the electron spins and
of the order of $\varepsilon_n$ for the
nuclear spins. The condition~(\ref{smallOmega}) can be written as
$B^1/(4\delta B)\ll 1$. If the distance between the qubits is 10 nm
and the magnetic field gradient is
$10^5$ T/m~\cite{gradient1,gradient2,gradient3}, then
$B^1\ll 4\delta B=2\times 10^{-3}$ T. Hence,
\begin{equation}
\label{smallOmega_e}
{\Omega_e\over 2\pi}\approx{\gamma_e\over 2\pi}B^1\ll 56 {\rm MHz},
\end{equation}

We now choose $K=K_e$ for the electron spin
to satisfy the following two conditions:
(i) The value of $K_e$ must be large enough
to satisfy the first equation~(\ref{smallOmega}), which allows one to 
decrease the error due to nonselective excitations; (ii) and the time $\tau_{e}$ 
of implementation of the Control Not gate on the (target) electron spin
must be much smaller than the electron relaxation time $T_c$
(we assume $T_c=60$ ms~\cite{T_c}). This condition can be satisfied
by decreasing $K_e$.

Consider the Control-Not gate on the electron spin with control nuclear spin.
The error $\sqrt{P_e}$
due to nonselective excitation of the electron spins is of the order of
[for $|\Delta|\approx A$ in Eq.~(\ref{2piK})]
\begin{equation}
\label{Error_e}
\sqrt{P_e}\sim{\Omega_e\over 4\gamma_e\delta B}\approx
{A\over 8\gamma_e\delta BK_e}\approx {1.03\over K_e},
\end{equation}
where we used the parameters
$A/2\pi=116$ MHz, the magnetic field
gradient $10^5$ T/m, and the distance 10 nm. In our simulations
we take $K_e=103$ so that the error is $\sqrt{P_e}\approx 0.01$. The
Rabi frequency and the time duration of the $\pi$-pulse are
$$
{\Omega_e\over 2\pi}\approx
{A\over 2\pi \sqrt{4K_e^2-1}}\approx 563~{\rm kHz},~~~
\tau_e\approx 0.89~\mu{\rm s}.
$$

Consider the Control-Not gate on the electron spin with control electron spin.
The error is of the order of ($|\Delta|\approx J$)
$\sqrt{P_e^\prime}\sim 0.01$ if
$K_e^\prime=1$, where we assume $J/(2\pi)=500$ kHz.
The Rabi frequency and the time duration of the $\pi$-pulse are
$$
{\Omega_e^\prime\over 2\pi}\approx
{J\over 2\pi \sqrt{4{K_e^{\prime 2}-1}}}\approx 289~{\rm kHz},~~~
\tau_e^\prime=2~\mu{\rm s}.
$$

Due to the second equation~(\ref{smallOmega}), the Rabi frequency
$\Omega_n$ must satisfy the condition
\begin{equation}
\label{smallOmega_n}
{\Omega_n\over 2\pi}\approx{2\gamma_n\over 2\pi}B^1\ll 34.5 {\rm kHz}.
\end{equation}
For implementation of the Control Not gate, $\Omega_n$ must also
satisfy the $2\pi K$ condition~(\ref{2piKn}).
The error in the Control-Not gate on the nuclear spin with control 
electron spin
is  $\sqrt{P_n}\sim 0.05$ if $K=K_n=33,620$.
The Rabi frequency and the time duration of the $\pi$-pulse acting
on a nuclear spin are
\begin{equation}
\label{Omega_tau_n}
{\Omega_n\over 2\pi}\approx
{A\over 2\pi \sqrt{4{K_n^{2}-1}}}\approx 1.7~{\rm kHz},~~~
\tau_n\approx 0.3~{\rm ms}.
\end{equation}

We note here that the magnetic field gradient
practically defines the clock speed of our quantum computer.
During the time of implementation of the Control Not
gate on the nuclear (target) spin, the
electron (control) spin must stay coherent so that the condition
$\tau_n\ll T_c$, where $T_c$ is the transverse relaxation time,
must be satisfied. For isotopically purified $^{28}$Si, the relaxation time
$T_c$ can be as long
as $60$ ms~\cite{T_c} at 7 K, which is large enough for implementation
of the Control-Not gate. In natural Si (4.7\% of $^{29}$Si) the
value of $T_c$ is smaller than 0.6 ms at
1.6 K~\cite{T_c}, which is too small for implementation
of the Control-Not gate on the nuclear spin. The Rabi frequency
of the order of 1.7 kHz is large in comparison to the
dipole-dipole interaction between the electron and nucleus of the
neighboring phosphorus atoms, which is close to 32 Hz for the distance 10 nm
between the qubits, so that one can neglect this dipole-dipole interaction.

We now make some comments about precision of calculation of
the energy levels in Section~\ref{sec:eigenstates}. As follows
from Eqs.~(\ref{solution}) and (\ref{solution1}), the spin rotates
around the $x$ axis (in the rotating frame) with the frequency
\begin{equation}
\label{lambda}
\lambda_{q,p}=\sqrt{\Delta_{q,p}^2+\Omega^2}.
\end{equation}
If we wish to implement
the resonant transition, $\Delta_{q,p}$ must be equal to zero. Since
$\Delta_{q,p}=E_q-E_p-\nu$ is defined by the distance between
the energy levels (eigenvalues), the complete transition takes place only if
these eigenvalues are exactly known. For a system with a small number
of qubits, the eigenvalues can be calculated numerically with a high
precision. For a system with a large number of qubits, ($>30$) one
can use the perturbation theory described in Section~\ref{sec:eigenstates}.
This theory allows one to calculate the eigenvalues
with precision $\epsilon^m A/2$, where
$\epsilon$ is defined in Eq.~(\ref{correction}) and $m$ is the order of the
perturbation theory ($m=2$ in our paper
and $m=1$ for the zeroth order approximation).
The limited accuracy of $E_q$ and $E_p$ results in a finite detuning of the
order of
$$
|\Delta_{q,p}|\sim \epsilon^m {A\over 2}.
$$
In order for this detuning to have a small influence on the dynamics,
the value of $|\Delta_{q,p}|$ in Eq.~(\ref{lambda}) must be much
smaller than the value of $\Omega$. From Eq.~(\ref{correction})
we have $\epsilon=6\times 10^{-4}$. In the zeroth order approximation
($m=1$), the value of $|\Delta_{q,p}/(2\pi)|$ is of the order of 35 
kHz, which is
much larger than $\Omega_n/(2\pi)$ in Eq.~(\ref{Omega_tau_n})
and is not acceptable. In the second order
approximation ($m=2$), we have $|\Delta_{q,p}/(2\pi)|\sim 21$ Hz, which is
much smaller than $\Omega_n/(2\pi)$ so that this is an
acceptable approximation for us. In practice, the above argument means
that the frequency $\nu$ of the electromagnetic wave must
be tuned with an accuracy of the order of several tens of Hertz
in order to flip a nuclear spin without generating a substantial error.

\subsection{Spin relaxation}
The relaxation of the electron spin can affect (flip) the nuclear spin
via the hyperfine interaction. Here we discuss the conditions required
to suppress the nuclear spin flip during the electron spin relaxation
process. Other kinds of relaxation mechanisms can
be neglected because the relaxation time $T_n$ of a neutron 
is at least four order of magnitude larger 
($T_n\sim 3\times 10^3$ s~\cite{relaxN}) than the total 
time of implementation of the algorithm.
The electron spin relaxation allows us to implement
the nonunitary transformation
\begin{equation}
\label{nonunitary}
\begin{array}{lll}
D_0|\Uparrow\downarrow\rangle & \rightarrow &
|\Downarrow\downarrow\rangle\\
& \nearrow & \\
D_1|\Downarrow \downarrow\rangle & &,
\end{array}
\end{equation}
which is necessary for creation of the initial state.

Consider the dynamics of the classical nuclear magnetic moment $\vec I(t)$
placed in a permanent external field $\vec B^0$ oriented along the z axis and
a field created by the electron spin $\vec S(t)$. Here the $z$ component
$S^z(t)$ of the electron spin is a given function of time $t$. The slowly
varying  $z$ component of the magnetic field acting on the nuclear spin is
\begin{equation}
\label{Bz}
B^z(t)=B^0-{A\over \gamma_n \hbar}S^z(t).
\end{equation}

The electron spin rotates with the frequency $\gamma_e B^0$
and generates a circularly polarized time-dependent
magnetic field acting on the nuclear spin via the hyperfine interaction.
Since $\gamma_e B^0$ is three orders of magnitude larger than the
Larmor frequency of the nuclear spin, this fast field
does not affect the dynamics of the nuclear spin
because the nuclear spin is out of resonance with the field generated
by the fast-rotating electron spin.

There are always stray magnetic fields in a real system,
such as, for example, the Earth's field,
which can affect the dynamics of the nuclear spin.
Without loss of generality, we assume that this field, $B^x$, is oriented
along the $x$ axis. The nuclear spin is flipped if
two conditions are satisfied. (a) The $B^z$ field (\ref{Bz})
must go through the zero point $B^z(t)=0$. (b) The condition of
adiabatic passage~\cite{adiabatic}
\begin{equation}
\label{adiabatic}
\xi={\left|\dot B^z\right|\over \gamma_n \left(B^x\right)^2}\ll 1
\end{equation}
must hold. Here $\xi$ is a dimensionless small parameter. 
We now analyze how to choose the parameters of our system
in order to violate these two conditions and, thus, to suppress the
nuclear spin flip.

Condition (a) is satisfied if the value of $B^z$
in Eq.~(\ref{Bz}) is always positive, which yields the minimum
value $B^0_{\rm min}$ of the external field (for $S^z=1/2$)
\begin{equation}
\label{adiabatic1}
B^0>B^0_{\rm min}={A\over 2\gamma_n \hbar}\approx 3.36~{\rm T}.
\end{equation}

Even if condition (a) is not satisfied, the nuclear spin cannot
flip if the magnetic field $B^x$ is sufficiently small to violate 
condition (b).
With a good approximation the function $S^z(t)$ can be chosen in the
form~\cite{Gorshkov}
\begin{equation}
\label{Szt}
S^z(t)={1\over 2}\left(1-2{t\over T_c}\right).
\end{equation}
The parameter $\xi$ in Eq.~(\ref{adiabatic}) becomes
\begin{equation}
\label{xi1}
\xi\approx{6.2\times 10^3\over \tilde T_c\left(\tilde B^x\right)^2},
\end{equation}
where the dimensionless relaxation time $\tilde T_c$ is equal to
the number of milliseconds [$\tilde T_c=T_c/(1~{\rm ms})$]
and the dimensionless $B^x$ field $\tilde B^x$ is equal to
the number of gauss [$\tilde B^x=B^x/(1~{\rm gauss})$].
We take $\tilde T_c=6$ (so that $T_c=6$ ms).
If $B^x$ is equal to the Earth's magnetic field, $\tilde B^x=0.5$,
the condition of adiabatic passage is not satisfied,
$\xi\approx 4.1\times 10^3\gg 1$
so that the nuclear spin does not flip.
The numerical modeling of the classical spin dynamics
with $B^x=0.5~{\rm gauss}$ and
\begin{equation}
\label{cl_parameters}
B^0=3.3T<B^0_{\rm min},~~~T_c=6~{\rm ms},~~~I^z(t=0)=-0.5
\end{equation}
yields
$$
\delta I^z={I^z(0)-I^z(T_c)\over I^z(0)}\approx 7.5\times 10^{-4}.
$$
One can show that the error in the quantum probability amplitude
is of the order of $\delta I^z$. One can neglect this error if it is small
in comparison with the other errors.
For $B^x=5~{\rm gauss}$ we have $\xi\approx 41$, and a
numerical simulation gives $\delta I^z\approx 0.075$.
For  $B^x=50~{\rm gauss}$ we have $\xi\approx 0.41$ and
numerical simulations show that the nuclear spin flips and
$I^z(T_c)\approx 0.5$. This is the situation when the conditions
of adiabatic passage are satisfied which prevent the initialization of
our computer.

If the transverse magnetic field is relatively strong, for example,
when $B^x=50~{\rm gauss}$, one can suppress the flip of the nuclear spin
by violating condition (a). For example, increasing $B^0$ from
$B^0=3.3$ T, which is less than $B^0_{\rm min}$ in Eq.~(\ref{adiabatic1}),
to $B^0=3.5~\rm{T}>B^0_{\rm min}$ and for the same transverse magnetic 
field $B^x=50~{\rm gauss}$, we numerically
obtained $\delta I^z(T_c)=7\times 10^{-4}$, i.e. the nuclear spin actually
does not flip.

\begin{figure}
%\vspace{-10mm}
\includegraphics[width=10cm,height=8cm]{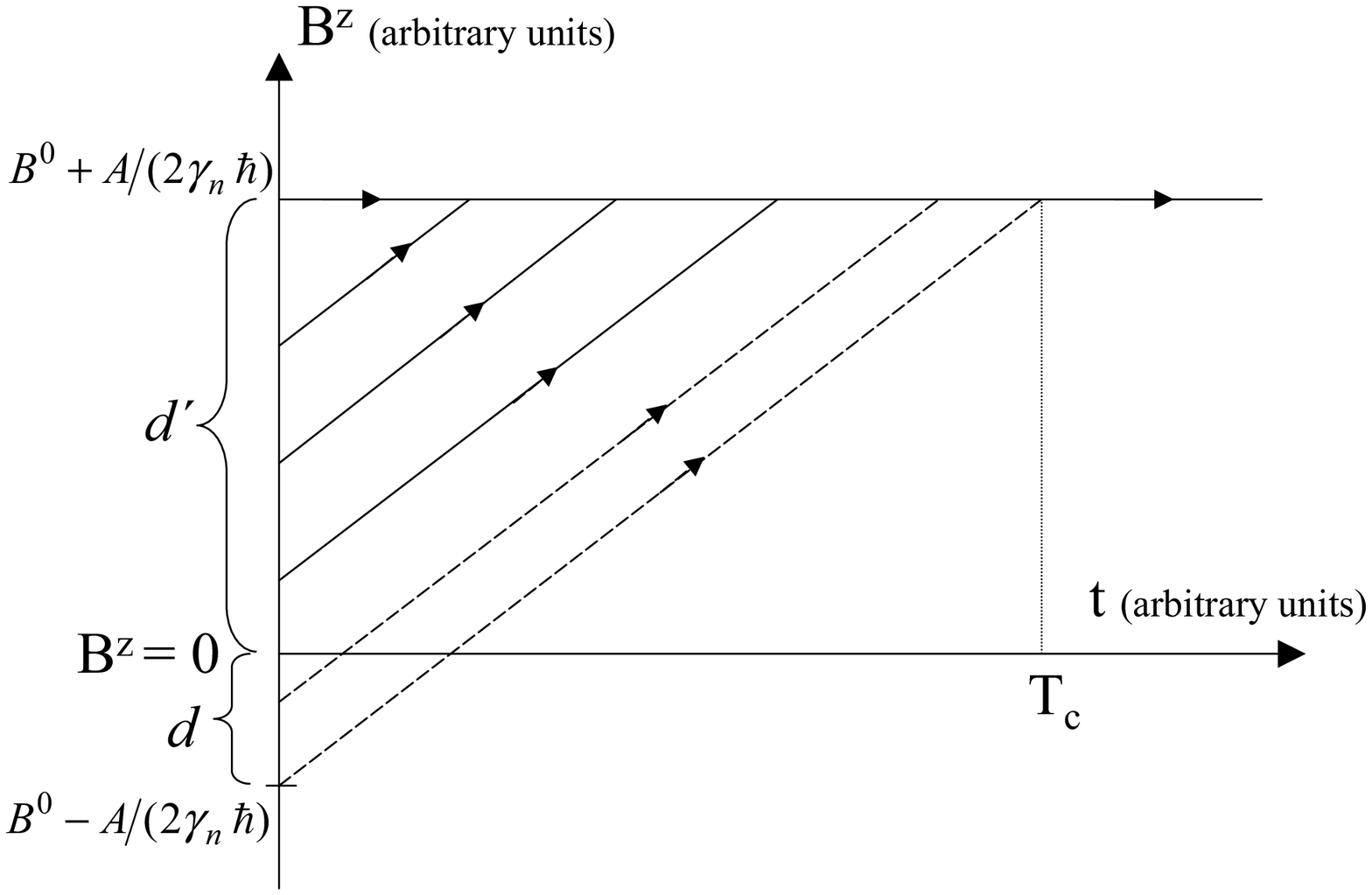}
\vspace{-5mm}
\caption{A schematic illustration of the magnetic field acting on
the nuclear spin for different values of $S^z(0)$
[$S^z(0)=1/2$ for the lowest curve and $S^z(0)=-1/2$
for the top curve]. The nuclear spin flips for the magnetic fields
illustrated by the dashed curves and does not flip for the
magnetic fields shown by the solid curves.}
\label{reffig:Szt}
\end{figure}

Next we will show that, in the situation when the conditions of
adiabatic passage are satisfied for certain spins, the error
still can be small if $B^0$ is close to $B^0_{\rm min}$, i.e. when
$|(B^0_{\rm min}-B^0)/B^0_{\rm min}|\ll1$.
Assume that we are dealing with an ensemble of identical spin chains
as mentioned in the introduction, $B^0<B^0_{\rm min}$,
and condition of adiabatic passage (\ref{adiabatic}) is satisfied.
Since before the relaxation
the electron spins point in a random direction, not all of them
pass the point for which $B^z(t)=0$ in Eq.~(\ref{Bz}).
In Fig.~\ref{reffig:Szt} we show how the magnetic field acting on
different nuclear spins changes with time. Only those
nuclear spins flip for which the condition
$$
B^0-{A\over 2\gamma_n\hbar}S^z(0)\le 0
$$
holds. As follows from the figure, the total number of such
spins in the ensemble is
\begin{equation}
\label{eta}
\eta={d\over d+d^\prime}=\frac 12-{\gamma_nB^0\hbar\over A}
=\frac 12\left(1-{B^0\over B^0_{\rm min}}\right).
\end{equation}
The error $\sqrt{P_{\rm r}}$
in the probability amplitude for the ensemble is $\sqrt{P_{\rm r}}\approx\eta$.
For example, for $B^0=3.3$ T (and for $B^0_{\rm min}=3.36$ T) we have
$\eta=0.009$ which is a small error.
In summary, our analysis shows that it is possible to suppress the flip
of the nuclear spin during the relaxation of the electron spin and
to implement the transformation~(\ref{nonunitary}).

\section{Numerical results}
\label{sec:numerical}
In our four-qubit system, all parameters of the pulses
can be calculated numerically using exact eigenvalues
of the matrix $\hat H^0$ and off-diagonal elements
of the matrix $\hat V$, which are equal to $\Omega_e/2$
for electron transitions and $-\Omega_n/2$ for the nuclear transitions.
In spite of the ability to calculate the parameters numerically,
we calculate them analytically using our perturbative approach
and simulate the quantum dynamics numerically as described in
Sec.~\ref{sec:numerical1}. Our analysis has the following advantages:
(i) it can be applied to a system with an arbitrary number of qubits
(ii) it allows one to take into consideration only
``slow'' transitions with small detunings and to neglect
fast transitions with relatively large detunings,
which have little influence the quantum dynamics.
Using our approach it is possible to understand the most important
sources of error and to minimize them by the optimal choice
of pulse parameters.

We start with the state
\begin{equation}
\label{start}
|\Psi(0)\rangle=C_6(0)|6\rangle+C_7(0)|7\rangle+C_{14}(0)|14\rangle
+C_{15}(0)|15\rangle
\end{equation}
with arbitrarily chosen complex coefficients $C_6(0)$, $C_7(0)$, $C_{14}(0)$,
and $C_{15}(0)$ at time $t=0$.
Then we make the transformation to the representation
of the Hamiltonian $\hat H^{0}$ [see Eq.~(\ref{a_ni})]
\begin{equation}
\label{transformCA}
D_n(0)=\sum_{i=0}^{15}a_{n,i}C_i(0).
\end{equation}
After initialization of the system and creation of entanglement,
we make the back transformation
\begin{equation}
\label{transformAC}
C_i(T)=\sum_{n=0}^{15}a_{i,n}D_n(T)e^{-iE_nT},
\end{equation}
where $T$ is the total time of implementation of the protocol.
The error is calculated as
\begin{equation}
\label{error}
P=\left|{1\over 2}-|C_0(T)|^2\right|+\left|{1\over 2}-|C_{15}(T)|^2\right|.
\end{equation}

\begin{figure}
%\vspace{-10mm}
\includegraphics[width=11cm,height=9cm]{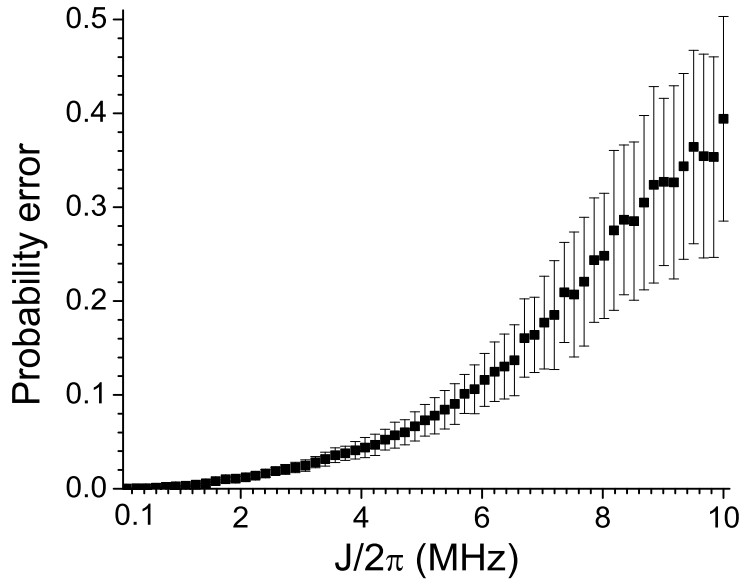}
\vspace{-10mm}
\caption{The probability error $P$ as a function of $J/(2\pi)$;
$\delta B=0.5$ mT.}
\label{reffig:PJ}
\end{figure}

In Fig.~\ref{reffig:PJ} we plot the error after implementation of
initialization and entanglement. Each point on the plot is the average
over 100 realizations with randomly chosen complex
coefficients $C_6(0)$, $C_7(0)$, $C_{14}(0)$, and $C_{15}(0)$.
One can see that the error increases with $J$ increasing.
When $J$ is large enough, $J\sim 4\gamma_e\delta B$, the basis
states $|n\rangle$ differ considerably from the eigenstates $|\psi_n\rangle$
of the Hamiltonian $\hat H^0$, so that the error is generated as a result
of free evolution of the basis states.
The error bars are the consequence of the fact that
the protocol processes some initial states of the superposition
better than other states.

\begin{figure}
%\vspace{-10mm}
\includegraphics[width=11cm,height=9cm]{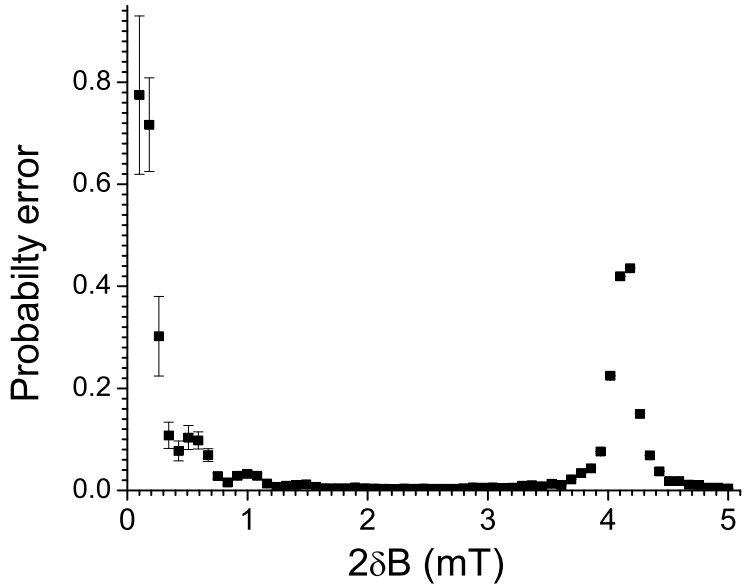}
\vspace{-10mm}
\caption{The probability error $P$ as a function of magnetic field difference
2$\delta B$ for $J/(2\pi)=2$ MHz. The data are averaged over 100 realizations
with randomly chosen normalized complex coefficients $C_6(0)$,
$C_7(0)$, $C_{14}(0)$, and $C_{15}(0)$.}
\label{reffig:PdB}
\end{figure}

In Fig.~\ref{reffig:PdB} we plot the probability error $P$ as a function
of the magnetic field difference $2\delta B=B_2^z-B_1^z$ for the
interval $0.1~{\rm mT}\le 2\delta B\le 5~{\rm mT}$. For a qubit 
spacing of 10 nm, this interval corresponds to
magnetic field gradients from $10^{-4}$ to
$5\times 10^{-5}$ T/m. As follows from the figure, the error is
large when $2\delta B$ is small, i.e. when $J/(4\gamma_e\delta B)\sim 1$.

The maximum in $P$ near $2\delta B\approx 4.2$ mT is defined
by the condition $\gamma_e\delta B =A/2$. When this condition is
satisfied the 10th and 12th eigenvectors defined in Eq.~(\ref{psi10})
and (\ref{psi12}) become symmetric and antisymmetric superpositions
$$
|\nnd\rangle\otimes\left[{1\over \sqrt 2}\left(
|\eed\eu\rangle\pm |\eeu\ed\rangle\right)\right]\otimes|\nuu\rangle.
$$
These states are formed because
the electron frequency difference caused by
the magnetic field gradient is compensated by the
hyperfine interactions between the electrons and nuclei.

The numerical results indicate that it is worthwhile to place
neighboring qubits at larger distance $d$ from each other.
This gives one the following advantages.
(i) At a given gradient, the value of $\delta B=(B^2_z-B^1_z)/2$ increases
with increasing $d$, which provides better selectivity of the pulses
and increases the clock speed of the quantum computer
[see the second equation~(\ref{smallOmega})
and Eq.~(\ref{Omega_tau_n})].
(ii) The value of $J$ decreases with increasing $d$ ~\cite{sarma}.
This does not affect the clock speed of the quantum computer
because the clock speed is defined by $\delta B$.
(iii) Decreasing $J$ decreases
the influence of the off-diagonal components of the
exchange interaction and the eigenstates $|\psi_n\rangle$ of the
Hamiltonian $\hat H^0$ are
better approximated by the basis states $|n\rangle$. Since
$|\psi_n\rangle\ne |n\rangle$, the error in the probability amplitude
[proportional to $J/(4\gamma_e\delta B$)] is generated
even in the stationary system when no electromagnetic pulses are applied.
(iv) In a system with more than two qubits
increasing $d$ decreases unwanted effect of long-range
interaction between distant (not neighboring) qubits. 
(v) When the distance between the qubits is large, the error is
less sensitive to the random qubit displacements caused by imperfect
qubit positioning using a scanning tunneling
microscope~\cite{to_be_published}.

\section{Summary}
We described how to implement quantum logic
operations in a silicon-based quantum computer with phosphorus atoms
serving as qubits. The logic operations can be implemented in our computer
if the following conditions are satisfied.
\begin{enumerate}
\item The selective excitations of nuclear spins can be implemented if
their Rabi frequencies are small [see the second equation~(\ref{smallOmega})],
\begin{equation}
\label{condition2000}
\Omega_n\ll 2\gamma_n(B_2^z-B_1^z),
\end{equation}
i.e., when $B^1\ll B_2^z-B_1^z$, where $B^1$ is the amplitude of the
radio-frequency field and $B_2^z-B_1^z$ is the magnetic field difference
equal to the product of the magnetic field gradient and the distance 
between the
qubits. Condition~(\ref{condition2000}) defines the clock speed of our
computer.  The Control Not gate between the nuclear spins is implemented
during the time-interval approximately equal to
$\tau_{CN}\approx\pi/\Omega_n$. The time-interval required to flip
the electron spins is at least two orders of magnitude smaller
than $\tau_{CN}$. As follows from Eq.~(\ref{condition2000}),
the clock speed is practically defined by the magnetic field gradient:
the larger is the gradient, the faster is the computer.
\item The time-interval $\tau_{CN}$ must be much smaller than
the electron relaxation time $T_2$ because the electron spins
should stay coherent during implementation of the Control Not gate.
\item In order to make the electron spins polarized,
the magnetic field $B^z$ must be large and the temperature $\Theta$
must be small, i.e., the condition $\gamma_e B^z\gg k_{\rm B}\Theta$
($k_{\rm B}$ is the Boltzmann constant) must be satisfied.
\item The electron-electron exchange interaction must be
small in comparison with the frequency difference
$\gamma_e(B_2^z-B_1^z)$ between the electrons, i.e. the
condition $\gamma_e(B_2^z-B_1^z)\gg J$ must be satisfied, otherwise, 
the off-diagonal components of the exchange interaction
would modify considerably the basis states and generate error.
\item The modified inequality $|\gamma_e(B_2^z-B_1^z)-A|\gg J$,
which includes the hyperfine interaction constant $A$,
must hold also.
\item In order to suppress the flip of the nuclear
spins during the relaxation of the electron spins,
the computer must be shielded from external stray
transverse magnetic fields, $B^x$, so that $B^x\le 5$ gauss or
the external permanent magnetic field $B^0$ must be close to
or larger than $B^0_{\rm min}=3.36$ T.
\end{enumerate}

\section*{Acknowledgments}
This work  was supported by
the Department of Energy under Contract No. W-7405-ENG-36 and DOE
Office of Basic Energy Sciences, by the National Security
Agency (NSA), and by Advanced Research and Development Activity (ARDA)
under Army Research Office (ARO) contract No. 707003.

\appendix
\section{}
Here we calculate the corrections $E^{(2)}_i$ to the eigenvalues
$E^{(0)}_i$ for some states using perturbation
theory~\cite{PRA01,JAM,book1}. The corrections to the
other eigenvalues are calculated in a similar fashion. The second order
correction $E^{(2)}_7$ to the eigenvalue $E^{(0)}_7$ is
\begin{equation}
\label{E27}
E^{(2)}_7=
\sum_n{|\langle\psi_7|\hat H^{(2)}|\psi_n\rangle|^2\over E^{(0)}_7-E^{(0)}_n}.
\end{equation}
The matrix elements of the
operator $\hat H^{(2)}$ are
[see Eq.~(\ref{HSigma-1})]
\begin{equation}
\label{H2Sigma-1}
\langle i|\hat H^{(2)}|j\rangle=\left(\begin{array}{cccc}
0 & 0 & {A\over 2} & 0 \\
0 & 0 & 0 & {A\over 2} \\
{A\over 2} & 0 & 0 & 0 \\
0 & {A\over 2} & 0 & 0
\end{array}\right).
\end{equation}
The basis vectors $|j\rangle$ are defined in Eq.~(\ref{psiSigma-1}).

The transformation from the eigenfunctions
$|\psi_n\rangle$ to the basis vectors $|i\rangle$ is described by the
matrix $a_{n,i}$ in Eq.~(\ref{a_ni}). We have
\begin{equation}
\label{7Vn}
\langle\psi_7|\hat H^{(2)}|\psi_n\rangle=
\sum_{i,j}a_{i,7}a_{n,j}\langle i|\hat H^{(2)}|j\rangle.
\end{equation}
 From equation~(\ref{E7}) we have $a_{i,7}=\delta_{i,7}$ where
$\delta_{i,j}$ is the Kronecker delta-function, and Eq.~(\ref{7Vn}) becomes
\begin{equation}
\label{7Vnn}
\langle\psi_7|\hat H^{(2)}|\psi_n\rangle=
\sum_{j}a_{n,j}\langle 7|\hat H^{(2)}|j\rangle.
\end{equation}
 From Eqs.~(\ref{psiSigma-1}) and (\ref{H2Sigma-1}), the only
nonzero matrix element is
\begin{equation}
\label{appen}
\langle 7|H^{(2)}|11\rangle={A\over 2}.
\end{equation}
 From Eqs.~(\ref{E27}), (\ref{7Vnn}), and  (\ref{appen}), we obtain
\begin{equation}
\label{E277}
E^{(2)}_7={A^2\over 4}{|a_{11,11}|^2\over E^{(0)}_{7}-E^{(0)}_{11}}.
\end{equation}
For our range of parameters, $|a_{11,11}|^2\approx 1$.
Using Eqs.~(\ref{E7}), (\ref{E11}), and (\ref{E277}),
we obtain Eq.~(\ref{E7(2)}).

Next we will calculate the correction $E^{(2)}_{13}$.
$$
E^{(2)}_{13}=\sum_n
{|\langle\psi_{13}|\hat H^{(2)}|\psi_n\rangle|^2\over E^{(0)}_{13}-E^{(0)}_n}.
$$
The matrix elements are
$$
\langle\psi_{13}|\hat H^{(2)}|\psi_n\rangle=\sum_{i,j}a_{i,13}a_{n,j}
\langle i|\hat H^{(2)}|j\rangle=
$$
$$
{A\over 2}(a_{11,13}a_{n,7}+a_{13,13}a_{n,14})=
{A\over 2}(a_{11,13}\delta_{n,7}+a_{13,13}\delta_{n,14}).
$$
The second-order correction is
$$
E^{(2)}_{13}={A^2\over 4}\left[
{|a_{11,13}|^2\over E^{(0)}_{13}-E^{(0)}_{7}}+
{|a_{13,13}|^2\over E^{(0)}_{13}-E^{(0)}_{14}}\right]\approx
{A^2\over 4}{1\over E^{(0)}_{13}-E^{(0)}_{14}}.
$$

In order to calculate the correction to the eigenvalue $E^{(0)}_6$, we note
that the state $|6\rangle$ is related by the matrix elements $A/2$
to the states $|5\rangle$ and $|10\rangle$ [which can be obtained
from the state $|6\rangle$ by swapping the states of $i$th nuclear
and electron spins, $i=1,2$]. After a brief calculation,
one can obtain Eq.~(\ref{E6(2)}).

\section{}
The corrections of the second order $E^{(2)}_i$, $i=1,14$,
to the energy levels $E^{(0)}_i$ are
$$
E_1^{(2)}\approx {A^2\over 4}
{1\over (\gamma_e+\gamma_n)b-\gamma_n\delta B-
\sqrt{(\gamma_e\delta B)^2+J^2/4}+J/2},
$$
$$
E_2^{(2)}\approx {A^2\over 4}
{1\over -(\gamma_e+\gamma_n)b+\gamma_n\delta B+
\sqrt{(\gamma_e\delta B)^2+J^2/4}-J/2}
$$
$$
E_3^{(2)}\approx 0,
$$
$$
E_4^{(2)}\approx {A^2\over 4}
{1\over -(\gamma_e+\gamma_n)b-\gamma_n\delta B-
\sqrt{(\gamma_e\delta B)^2-J^2/4}+J/2},
$$
$$
E_5^{(2)}\approx {A^2\over 4}\left[
{1\over E_5^{(0)}-E_{6}^{(0)}}+{1\over E_5^{(0)}-E_{9}^{(0)}}\right],
$$
[the values $E_i^{(0)}$, $i=1,\dots,14$ are defined in
Section~\ref{sec:eigenstates}],
\begin{equation}
\label{E6(2)}
E_6^{(2)}\approx {A^2\over 4}\left[
{1\over E_6^{(0)}-E_{5}^{(0)}}+{1\over E_6^{(0)}-E_{10}^{(0)}}\right],
\end{equation}
\begin{equation}
\label{E7(2)}
E_7^{(2)}\approx {A^2\over 4}
{1\over -(\gamma_e+\gamma_n)b-\gamma_n\delta B-
\sqrt{(\gamma_e\delta B)^2+J^2/4}+J/2},
\end{equation}
$$
E_8^{(2)}\approx {A^2\over 4}
{1\over (\gamma_e+\gamma_n)b+\gamma_n\delta B+
\sqrt{(\gamma_e\delta B)^2+J^2/4}+J/2},
$$
$$
E_9^{(2)}\approx {A^2\over 4}\left[
{1\over E_9^{(0)}-E_{5}^{(0)}}+{1\over E_9^{(0)}-E_{10}^{(0)}}\right],
$$
$$
E_{10}^{(2)}\approx {A^2\over 4}\left[
{1\over E_{10}^{(0)}-E_{6}^{(0)}}+{1\over E_{10}^{(0)}-E_{9}^{(0)}}\right],
$$
$$
E_{11}^{(2)}\approx {A^2\over 4}
{1\over (\gamma_e+\gamma_n)b+\gamma_n\delta B+
\sqrt{(\gamma_e\delta B)^2+J^2/4}-J/2},
$$
$$
E_{12}^{(2)}\approx 0,
$$
$$
E_{13}^{(2)}\approx {A^2\over 4}
{1\over (\gamma_e+\gamma_n)b-\gamma_n\delta B-
\sqrt{(\gamma_e\delta B)^2+J^2/4}-J/2},
$$
$$
E_{14}^{(2)}\approx {A^2\over 4}
{1\over -(\gamma_e+\gamma_n)b+\gamma_n\delta B+
\sqrt{(\gamma_e\delta B)^2+J^2/4}+J/2},
$$
In the text we assume $E_i=E^{(0)}_i+E^{(2)}_i$.

\end{document}